# Chapter 34 - Every Moment Counts: Synchrophasors for Distribution Networks with Variable Resources

Alexandra von Meier, Reza Arghandeh

California Institute for Energy and Environment
University of California, Berkeley
Berkeley, CA 94794, USA

**Abstract**

Historically, with mostly radial power distribution and one-way power flow, it was only necessary to evaluate the envelope of design conditions, e.g., peak loads or fault currents, rather than continually observe the operating state. But the growth of distributed energy resources introduces variability, uncertainty, and opportunities to recruit diverse resources for grid services. This chapter addresses how the direct measurement of voltage phase angle might enable new strategies for managing distribution networks with diverse, active components.

## 1. Introduction

Historically, power distribution systems did not require elaborate monitoring schemes. With radial topology and one-way power flow, it was only necessary to evaluate the envelope of design conditions, i.e. peak loads or fault currents, rather than continually observe the operating state. But the growth of distributed energy resources such as renewable generation, electric vehicles and demand response programs introduces more short-term and unpredicted fluctuations and disturbances [1]. This suggests a need for more refined measurement, given the challenges of managing increased variability and uncertainty, and the opportunities to recruit diverse resources for services in a more flexible grid. This chapter addresses how the direct measurement of voltage phase angle might enable new strategies for managing distribution networks with diverse, active components. Specifically, it discusses high-precision synchrophasors, or micro-phasor measurement units (μPMUs), that are tailored to the particular requirements of power distribution in order to support a range of diagnostic and control applications, from solving known problems to opening as yet unexplored possibilities.

## 2. Variability, Uncertainty and Flexibility in Distribution Networks

Electric transmission and distribution systems are formally distinguished by voltage level, but harbor profound differences in design and operation. These differences explain the diverse sets of challenges encountered in the context of renewables integration, as well as the historical lag of distribution behind transmission systems in terms of observability and sophistication of measurement. Broadly speaking, distribution systems tend to be low-tech, aging, and due for upgrades[2-4].

*Architecture:* For economy and simplicity of protection, distribution systems are generally laid out radially, with legacy equipment such as protective relays and voltage regulation devices designed on the assumption of one-directional power flow from the substation toward loads. While easier to operate in principle, radial design also presents liabilities: When distributed generation introduces reverse power flow, some older controls could malfunction. Mathematically, radial design makes the estimation of the operating state more difficult, by removing the redundancy afforded by Kirchhoff's laws: in other words, the estimate for voltages and currents at one node cannot be corroborated by those at neighboring nodes. Distribution systems also have a much greater number of nodes or branch points than transmission networks, since each secondary transformer represents a load bus. With many more nodes than measuring points (and without smart meter data available in real-time), state estimation becomes even more difficult to perform. Traditional distribution operations never required this level of analysis, but the uncertainties introduced by diverse distributed resources make it increasingly important to assess the actual operating state of the system.

*Variation*: At the local scale, we lose the statistical benefit of aggregating large numbers of customers, as assumed in transmission level analysis. Consequently, there is more variation. The load duration curve is "peakier" for an individual distribution feeder than for a larger territory, and ramp rates as a percentage of load are much steeper. Also, phase imbalances matter (sometimes in the tens of percent), making it necessary to consider all three phases individually. Variation means that local idiosyncracies such as load types and topography are more important: no two distribution feeders are exactly alike, and it can be difficult to extrapolate analytical findings from one area to another. This underscores the need for carefully monitoring individual distribution circuits as penetration levels of active components increase.

*Exposure:* Closer proximity to many types of hazards – flora, fauna, and human activities – means more exposure and vulnerability for distribution circuits: unsurprisingly, a large majority of customer outages originate in the distribution system. With any number of local factors impacting distribution, but without redundant supply paths, distribution operations often revolve around switching procedures such as isolating sections or restoring service to customers as safely and quickly as possible. This suggests that improved distribution reliability is a likely area for early benefits from advanced monitoring. In general, exposure implies a higher degree of uncertainty in distribution operations, while operating errors and malfunctions pose a very immediate and physical risk.

*Opacity:* Historically, distribution operators have relied on field crews as their eyes and ears to report on system status. Despite increasing prevalence of supervisory control and data acquisition (SCADA), it is still often necessary to send someone in a truck to verify, for example, whether a switch is open or closed, or to pinpoint the location of a downed line. This has important practical implications, but also poses an analytical challenge: while the power flow calculation in transmission networks assumes that the topology of the network and the physical characteristics of all branches are known exactly, such information tends not to be reliably available for distribution circuits.

Various tools have been developed and implemented to provide distribution operations and planning with more detailed and timely information. Even so, creating situational awareness out of disjointed data streams remains a challenge. Circuit models, where available, may be based on unreliable input data and questionable assumptions. Physical measurements from the field remain a limiting factor for analysis, human operators and automated control systems alike.

The lack of visibility on distribution systems follows from simple economics: there has never been a pressing need to justify extensive investment in sensing equipment and communications. Even with the growing need for monitoring capabilities, the costs must be far lower to make a business case for measurement devices on a distribution circuit as compared to the transmission setting.

Arguably, the time for such a business case is fast approaching. Distributed energy resources are beginning to pose both challenges and opportunities for actively managing distribution circuits. As discussed elsewhere in this volume, renewable and other non-traditional resources create a need for coordination at higher resolution in both space and time, from protection to voltage regulation and other power quality issues. They also represent a new menu of options for grid support functions such as volt-VAR optimization, energy storage on time scales anywhere from cycles to hours, or even intentional islanding. Both to avoid adverse customer impacts and take optimal advantage of new resources, increased flexibility in operating distribution circuits is called for. But given the particular data-richness of distribution circuits, this means that much more, better and faster information from far behind the substation will be needed to make intelligent and economical decisions. This chapter suggests the possibility that distribution system monitoring and control might leapfrog a generation of SCADA developed for transmission systems to proceed straight into the 21$^{st}$ century, bringing us to the state of the art in a.c. measurement: synchrophasors.

### 3. Micro-Synchrophasor (µPMU) Technology

The essense of synchrophasors, or phasor measurement units (PMUs), is the precise time-stamping of voltage measurements to compare the phase angle between different locations [5]. This technology became feasible in the 1980s with readily accessible GPS time signals. Real (active) power flow between two points on an a.c. network varies mainly with the voltage angle difference δ. When the line impedance is mainly inductive, real power flow $P_{12}$ can be approximated by the scalar equation

$$P_{12} \approx \frac{V_2 V_1}{X} \sin \delta \tag{1}$$

where X is the line inductance and $V_1$ and $V_2$ are the voltage magnitudes. Voltage magnitude and phase angle at each node in a network are considered the *state variables*, since they uniquely determine a.c. power flow throughout the network.

Direct measurement of the state variable δ offers some basic advantages. Voltage angle can serve as a proxy for local current measurements where installation of current sensors is inconvenient. If both voltage and current magnitude and angle are measured, this provides maximal information about the system state from just one instrumented node. Because voltage angle varies across an a.c. network in a continuous profile, power flow patterns can be inferred from angle gradients, without explicitly measuring branch currents. Beyond steady-state analysis, the key benefit of synchrophasor measurements lies in observing dynamic behavior, including rapid changes on the scale of cycles rather than seconds.

Today, PMUs are used almost exclusively on transmission systems. Some of the most notable benefits to date have come from observing sub-synchronous oscillations across wide areas, such as the Western Interconnect in the U.S., that threaten a.c. system stability. By directly identifying oscillation modes and their associated damping, operators can take specific appropriate actions (such as de-rating transmission lines) as necessary, thus supporting both reliability and asset utilization.

Although "distribution PMUs" may already be deployed at distribution substations (for example, embedded in protective relays), the use of their measurements to date is mainly in reference against phase angles elsewhere on the transmission grid, not the distribution feeder. By contrast, the purpose of micro-synchrophasors is specifically to compare voltage angles at different points on distribution circuits, behind the substation. According to Equation (1), such angle differences on distribution systems will tend to be much smaller, owing to smaller power flows and shorter distances. For example, a typical voltage phase angle difference for a distribution feeder at full load might be 0.1° per mile, as compared to tens of degrees between transmission nodes.[1] Consequently, transmission PMUs with typical errors near ±1° may not provide enough precision for meaningful distribution measurements.

Besides requiring greater angular resolution, distribution synchrophasor measurements will likely have to contend with more noise, including harmonics and small transients associated with nearby devices or switching operations on a circuit. For this reason, we expect that it will prove useful to combine PMUs with power quality measurements, so as to analyze and interpret angle data in proper context.

Figure 1 illustrates the capabilities of a new μPMU device developed and manufactured by Power Standards Laboratory, based on a commercially available power quality recorder, the PQube (www.powerstandards.com). A key capability of the μPMU is to combine high-resolution angle measurements with detailed characterization of waveforms, including harmonics and transients. The figure shows relevant quantities are approximately situated on a logarithmic time scale for visual comparison . An absolute floor of attainable angular resolution is set by timekeeping accuracy, while the high sampling rate serves to characterize harmonics. Useful rates of data recording and communication will vary depending on what practical applications in distribution systems are to be supported, which will differ in the quantity, quality, and timeliness of measurement data they require.

---

[1] It should be noted that the approximate Equation (1) is less apt when applied to distribution lines, where resistance is significant compared to inductance (greater R/X ratio). However, the general statement holds that phase angle differences in distribution are one to two orders of magnitude smaller than in transmission.

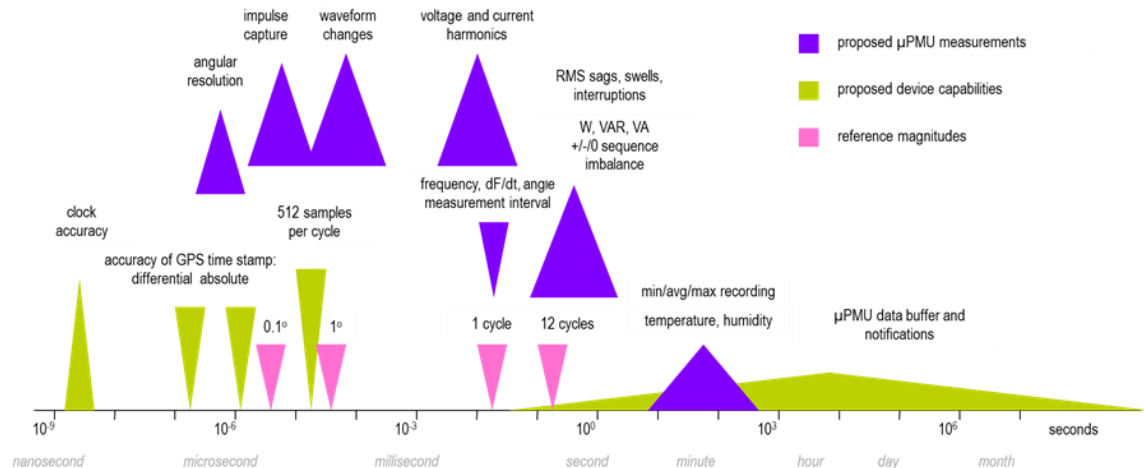

**Figure 1. Time scale of μPMU functionalities**

The authors are using this μPMU in a project funded by the U.S. Department of Energy's ARPA-E program.[2] We envision from several up to tens of μPMUs installed at multiple locations throughout a distribution feeder – for example, at the substation, end of feeder or laterals, and any key distributed generation facilities. The μPMU connects at the secondary voltage level (with voltage inputs from 100~690$V_{LL}$) or through potential transformers at substations or line devices. Each μPMU may upload its precisely time-stamped measurements through a suitable communication layer (for example, 4G wireless) to a flexible local network where angle measurements will be compared. The network concept will build on the simple Measurement and Actuation Profile (sMAP) developed by UC Berkeley as a foundation for managing both real-time and archival data from a wide variety of physical sources [6]. Initial μPMU deployment schemes and networking will be tested in several field installations under the ARPA-E project[7].

## 4. Applications for μPMU Measurements

A broad spectrum of potential distribution system applications could hypothetically be supported by synchrophasor data, as has been noted in the literature [8-10]. This section aims to characterize some selected applications and speculate about the potential advantage afforded by voltage angle measurement to support them.

---

[2] This three-year award, DOE 0000340, commenced in 2013 and includes research partners CIEE, UC Berkeley, Lawrence Berkeley National Laboratory, and Power Standards Lab.

*Topology status verification* means detecting or confirming the open or closed status of operable switches or breakers by comparing voltage angle directly across them or validating power flow solutions under different network topologies. Fast and reliable empirical topology identification could enhance safety where remote indicators are unavailable or considered unreliable. It could also support circuit switching operations for various purposes, including service restoration after an outage, or safely accommodating certain microgrid configurations.

*Unintentional Island Detection* is a special case of topology detection. The goal is to immediately identify when distributed generators continue to energize a local portion of the network that has been separated from the main grid, and to reliably distinguish dangerous fault situations from other abnormal conditions where it may be desirable to keep DG online.

*Phase identification and balancing* is another special case of topology detection. Direct angle measurement with a portable µPMU on the secondary distribution system would be a uniquely quick and easy way to ascertain which single-phase loads are connected to A, B or C. Improved phase balancing can reduce losses and increase the overall efficiency of operating a three-phase system.

*State estimation* means identifying as closely as possible, from available network models and empirical measurements, the operating state of the a.c. system in near real-time – that is, to specify voltage magnitude and phase angle for every node in the network. Data from µPMUs could ease the difficulties of distribution state estimation by directly feeding state variables into a Distributed State Estimator, which in turn may provide information to a Distribution Management System (DMS).

*Reverse power flow* is a special case of an operating state that might be undesirable. Phase angle measurements from suitable locations across a distribution circuit may help anticipate when and where reverse power flow is likely to occur.

*Fault location* is a critical function for both safety and speed of restoration. The goal is to infer the actual geographical location of a fault on a distribution feeder to within a small circuit section between protective devices, by comparing recorded measurements of voltage angle before and during the fault and interpreting these in the context of a circuit model. While various fault location algorithms exist, the quality of available measurements on distribution circuits are often insufficient to support them. Voltage angle at points some distance away from a fault may prove to be a more sensitive indicator of fault location than voltage magnitude.

*High-impedance fault detection* means recognizing the dangerous condition where an object such as a downed power line makes an unintentional connection with the ground, but does not draw sufficient current to trip a protective device since it mimics a legitimate load. If a combination of high-resolution angle and power quality measurements could help distinguish high-impedance faults from loads, this would afford considerable safety benefits.

*Dynamic circuit characterization* involves observing and studying the behavior of distribution circuits on very short time scales that has not been readily observable to date. Dynamic behavior in the presence of high penetrations of distributed resources may or may not involve anything problematic or actionable; the point is that we don't presently know and it may be worth looking.

*Oscillation detection* is one aspect of dynamic analysis. PMUs have shown subsynchronous oscillations to exist on transmission systems (where they indeed caused problems) that were neither predicted by models nor observed by conventional instrumentation. Higher-frequency oscillations could conceivably occur on distribution systems as a result of power exchange between and among distributed energy resources, or any resonance phenomena on the circuit [11, 12]. If so, µPMU measurements would be an ideally suited diagnostic tool[13].

*Characterization of distributed generation* could be performed with µPMU measurements at very small time scales below a cycle. The goal is to qualify and quantify the behavior of inverters in relation to stabilizing system a.c. frequency and damping disturbances in power angle or frequency. This could help rule out adverse grid impacts such as resonance or simultaneous trips of distributed generators, while facilitating their recruitment for advanced ancillary services such as inertia or transient mitigation in the future.

*Unmasking loads from net metered DG* would involve inferring the amount of load being offset by DG behind a net meter through measurements and correlated data obtained outside the customer's premises. Estimating the real-time levels of renewable generation versus loads would allow for better anticipation of changes in the net load, by separately forecasting the load and generation, and for assessing the system's risk exposure to sudden generation loss. At the aggregate level, this information is of interest to system operators for evaluating stability margins and damping levels in the system.

*Fault-induced delayed voltage recovery (FIDVR)* is an unstable operating condition that results from the interaction of stalled air conditioners with capacitor bank controls [14]. μPMU data might help anticipate FIDVR before it occurs, by identifying in near real-time the varying contribution to total customer load from devices such as single-phase induction motors in residential and small commercial air conditioners that pose an increased risk.

Table 1 summarizes prospective advantages of high-resolution voltage angle measurements as compared to conventional techniques to support the listed diagnostic applications.

Table 1. Potential diagnostic applications with μPMU measurements

| Diagnostic Application | Competing conventional strategies | Likely advantage of voltage angle | Likely technical challenges |
|---|---|---|---|
| **Topology detection** | direct SCADA on switches | possibly fewer measurement points, independent validation | algorithm using minimal placement |
| **Unintentional island detection** | various | possibly faster, greater sensitivity and selectivity, possibly less expensive | speed |
| **State estimation** | computation based on V mag measurements | possibly fewer measurement points, better accuracy, faster convergence | algorithm using minimal placement |
| **Reverse power flow detection** | detect with PQ sensor (V mag, I mag & angle) | may extrapolate to locations not directly monitored | algorithm using minimal placement |
| **Fault location** | various | possibly better accuracy, i.e. locate fault more closely with δ than V | need high resolution, fast data |
| **High-impedance fault detection** | various, difficult | possibly better sensitivity and selectivity with δ | unknown |
| **Dynamic circuit monitoring** | high-resolution PQ instruments, none for δ | uniquely capture oscillations, damping | data mining for relevant phenomena |
| **Oscillation detection** | none | unique | unknown |
| **Load and DG characterization** | limited observation with PQ instruments | uniquely capture dynamic behaviors | data mining, proximity to subject |
| **Unmasking load/DG** | none | may be unique | unknown |
| **FIDVR detection** | detected with V mag | possibly less expensive, faster | easy |
| **FIDVR prediction** | none | may be unique | unknown |

The above applications focus on diagnostic capabilities. However, synchrophasor data may enable more refined management and active control of distribution systems. Possible control applications include the following:

*Protective Relaying.* Reverse power flow was noted above as a condition that can be important to diagnose and avoid, but another approach is to employ protection schemes that safely accommodate reverse flow. Without requiring a costly replacement of protective devices, it may be feasible to develop supervisory differential relaying schemes based on μPMU data that recommend settings to individual devices according to overall system conditions [15].

*Microgrid Coordination.* Microgrid monitoring and control has to satisfy a number of objectives: load sharing among DG units, voltage and frequency regulation in islanded and grid connected modes, island detection and resynchronization, supply and demand optimization, and real-time monitoring of disturbances and harmonics. As an accurate measure of system state, μPMU data could support control algorithms in all of the above aspects.

Generation and load within a power island can be balanced through conventional frequency regulation techniques, but explicit phase angle measurement may prove to be a more versatile indicator. In particular, angle data may provide for more robust and flexible islanding and re-synchronization of microgrids. A convenient property of PMU data for matching frequency and phase angle is that the measurements on either side need not be at the identical location as the physical switch between the island and the grid. A self-synchronizing island that matches its voltage phase angle to the core grid could be arbitrarily disconnected or paralleled, without interruption of load. Initial tests of such a strategy with angle-based control of a single generator were found to enable smooth transitions under continuous load with minimal discernible transient effects [16].

Comparison of angle difference between a microgrid or local resource cluster and a suitably chosen point on the core grid could enable the cluster to provide ancillary services as needed, and as determined by direct, physical measurement of system stress rather than a price signal – for example, by adjusting power imports or exports to keep the phase angle difference within a predetermined limit. A variation of this approach, known as angle-constrained active management (ACAM), has been demonstrated in a limited setting with two wind generators on a radial distribution circuit [17].

*Volt-VAR Optimization*. Voltage angle measurement would not afford an inherent advantage over magnitude for feeder voltage optimization, but the capability to support this important function alongside other applications could add significantly to the business case for μPnet deployment.

## 7. Moving Forward

This chapter proposed high-resolution voltage phase angle measurement as a new option for accurate and flexible monitoring and control of distribution networks in the presence of variability and uncertainty. The central hypothesis is that a broad range of specific diagnostic and control applications will depend on improved visibility and transparency of the distribution system, meaning better knowlegde of the system state in real-time. That is, to effectively manage distribution networks with high renewable penetration, demand response and distributed control, high-precision monitoring systems will be needed to provide clear, accurate and complete observation of varying system behavior. The authors believe that μPMUs are a strong candidate for creating this functionality in an economical manner.

Before any of the specific applications discussed here can be evaluated in practice, it will be necessary to simply observe what phenomena can in fact be detected at the resolution of the μPMU, and what can be reliably concluded from those observations. A key challenge will be to distill raw phase angle measurements into tools that support situational awareness and ultimately produce actionable operational intelligence, without the clutter of excess data. Also, to advance opportunities for active coordinated control based on μPMU measurements, the requirements for hierarchical, layered, distributed control of aggregated distributed resources and loads – in particular, clusters capable of islanding – need to be studied in relation to voltage phase angle.

Active management of transparent distribution systems at high granularity in space and time may become a necessity simply to accommodate new resources without adverse impacts on power quality and reliability. It also implies exciting possibilities for new operating strategies: for example, distributed resources might smoothly transition between connected and islanded states, and be capable of providing local power quality and reliability services on the one hand and support services to the core grid on the other, as desired at any given time. This type of flexibility is essentially a form of redundancy, without a clear business case at present. However, considerations of security and infrastructure resiliency may support the development of such strategies in the future. Whatever course the evolution of distribution systems takes, increased visibility and precise measurement will be critical aspects of the new infrastructure.